\documentstyle[12pt]{article}
\begin{document}

\title{Doubly Lopsided Mass Matrices from Unitary Unification}

\author{{\bf S.M. Barr} \\
Bartol Research Institute \\ University of Delaware \\
Newark, Delaware 19716}

\date{\today}

\maketitle

\begin{abstract}
It is shown that the stratified or ``doubly lopsided" mass matrix structure
that is known to reproduce well the qualitative features of the quark and
lepton masses and mixings can arise quite naturally in the context of grand unification
based on the groups $SU(N)$ with $N > 5$.  An $SU(8)$ example is constructed with the minimal
anomaly-free, three-family set of fermions, in which a
realistic flavor structure results without
flavor symmetry.
\end{abstract}

\newpage

\section{Introduction}

A still unanswered question is why the quarks and leptons of
different families have different masses even though they transform
in exactly the same way under the symmetries of the Standard Model.
Most proposed answers are based on the idea that there are flavor
symmetries that distinguish fermions of different families. There is
another idea, however, suggested long ago \cite{barr80} but much less studied,
which is that there is a grand unified gauge group, $G$,
under which different families transform differently. If $G = SU(N)$, then
$N$ must be greater than 5, since under $SU(5)$ every
family transforms the same way, namely as
${\bf 10} + \overline{{\bf 5}}$. Under $SU(N)$, with
$N>5$, however, families or parts of families can come from multiplets
of various sizes.

For instance, consider $SU(6)$ with fermion multiplets
that include totally antisymmetric rank-2 and rank-3 tensors:
$\psi^{AB} = {\bf 15}$ and $\psi^{ABC} = {\bf 20}$. Both the
${\bf 15}$ and the ${\bf 20}$ contain a ${\bf 10}$ of $SU(5)$ and
therefore contain fermions with the quantum numbers of $u_L$, $d_L$,
$u^c_L$, and $e^+_L$.
Suppose further that the weak-interactions were broken only by
a Higgs field that is in a ${\bf 15}$ of $SU(6)$. Then the only mass
term for the up-type quarks allowed by $SU(6)$ would be of the form
$\psi^{AB} \psi^{CD} \langle H^{EF}
\rangle \epsilon_{ABCDEF}$, i.e. ${\bf 15} \; {\bf 15} \; \langle {\bf 15}_H \rangle$,
which gives mass only to the up-type quark in the ${\bf 15}$, but
not to the up-type quark in the ${\bf 20}$. Therefore, without any ``flavor
symmetry", a hierarchy of fermion masses would result. ($SU(6)$ is not
large enough to give interesting or realistic examples; but
simple realistic examples can be constructed with $SU(N)$ groups
with $N\geq 7$. A realistic $SU(8)$ example will be presented below. For models
implementing a similar ``flavor without
flavor symmetries" idea using the group $SO(10)$, see \cite{flavwoflav}.)

There are several ways that hierarchies can arise among the
light fermion masses in such schemes. In a fermion mass matrix, some
elements may arise from renormalizable Yukawa terms (like the
${\bf 15} \; {\bf 15} \; {\bf 15}_H$
term in the $SU(6)$ example), some may arise from higher-dimension
operators generated by tree diagrams, and some may arise from
higher-dimension operators generated by loop diagrams. Even elements
that arise from operators of the same dimension and at the
same loop level can still have very different magnitudes if
the operators that produce them involve Higgs
fields that transform differently under $G$.

In $SU(N)$ with the normal embedding of the Standard Model
group, there are no exotic fermions if all the fermion
multiplets are totally antisymmetric tensors.  A rank-$p$
totally antisymmetric tensor will be denoted by
$[p]$ and its conjugate tensor by $\overline{[p]}$ or by $[N-p]$.
If the set
of fermions multiplets is anomaly-free, then, as is well-known,
they decompose under the $SU(5)$ subgroup as some number
of ${\bf 10} + \overline{{\bf 5}}$ families together with
a vectorlike set of multiplets that can
contain ${\bf 10} + \overline{{\bf 10}}$ pairs,
${\bf 5} + \overline{{\bf 5}}$ pairs, and singlets. As
there is typically no symmetry to prevent it, the conjugate pairs in the
vectorlike set ``mate" with each other to acquire superheavy mass.
The ${\bf 10} + \overline{{\bf 5}}$ families, however, being chiral,
are forbidden to obtain mass and remain light. (This is Georgi's well-known
``survival hypothesis" \cite{survhypoth}.)
Therefore, the fact that the observed light fermions
fit neatly into some number of ${\bf 10} + \overline{{\bf 5}}$
families of $SU(5)$, which is often seen as pointing to $SO(10)$
unification, has just as simple an explanation
in terms of $SU(N)$ unification.  Moreover, $SU(N)$ has the following
theoretical advantage over $SO(10)$: In $SO(10)$ the simplest possibility
is that all the ${\bf 10} +\overline{{\bf 5}}$ come from ${\bf 16}$
spinor multiplets, so that the gauge group does not distinguish among
the families. But for $SU(N)$, as we will see in the
$SU(8)$ example described below, it can happen that even with the
{\it simplest} anomaly-free three-family set of fermion multiplets,
the three light families do not transform in the same way under
the $SU(N)$ group.

Before describing what happens in $SU(N)$, it will be useful to set the
stage by reviewing some recent ideas for explaining the gross features of
the observed patterns of quark and lepton masses and mixings in the context
of $SU(5)$.  It will be seen below that the $SU(5)$ structures postulated
by these recent ideas emerge automatically in $SU(N)$ unification.

The recent $SU(5)$-based idea is that of ``doubly lopsided" mass
matrices. (The first paper proposing the lopsided mass matrix idea
\cite{bb1996} actually proposed the doubly lopsided structure.
Singly lopsided --- or just ``lopsided" --- models were
independently proposed by several groups to explain the large
atmospheric neutrino mixing angle \cite{lopsided}. For a review see
\cite{lopsidedreview}. Then doubly lopsided models were taken up
again by several groups as an explanation of the fact that both the
atmospheric and solar angles are large \cite{hm2001,bb2002}.) The
doubly lopsided structure emerges naturally as follows.

Imagine that some symmetry distinguishes the three
light ${\bf 10}$'s of quarks and leptons and prevents them from mixing
strongly with each other. Let the mixing of ${\bf 10}_1$ with ${\bf 10}_2$
be controlled by the small parameter $\delta$ and the mixing of
${\bf 10}_2$ with ${\bf 10}_3$
be controlled by the small parameter $\epsilon$. On the other hand, imagine
that no symmetry distinguishes the light
$\overline{{\bf 5}}$'s from each other, so that they are allowed to mix strongly.
In that case one would expect the following structures for the three types of mass
matrices (the entries in the matrices give only the order of magnitude of
the elements):

\begin{equation}
\begin{array}{l}
({\bf 10}_1, {\bf 10}_2, {\bf 10}_3) \left( \begin{array}{ccc}
\delta^2 \epsilon^2 & \delta \epsilon^2 & \delta \epsilon \\
\delta \epsilon^2 & \epsilon^2 & \epsilon \\
\delta \epsilon & \epsilon & 1
\end{array} \right) \left(
\begin{array}{c} {\bf 10}_1 \\ {\bf 10}_2 \\ {\bf 10}_3 \end{array}
\right) \langle {\bf 5}_H \rangle, \\ \\
({\bf 10}_1, {\bf 10}_2, {\bf 10}_3) \left( \begin{array}{ccc}
\delta \epsilon & \delta \epsilon & \delta \epsilon \\
\epsilon & \epsilon & \epsilon \\
1 & 1 & 1
\end{array} \right) \left(
\begin{array}{c} \overline{{\bf 5}}_1 \\ \overline{{\bf 5}}_2 \\
\\ \overline{{\bf 5}}_3 \end{array}
\right) \langle \overline{{\bf 5}}_H \rangle, \\
(\overline{{\bf 5}}_1, \overline{{\bf 5}}_2, \overline{{\bf 5}}_3)
\left( \begin{array}{ccc}
1 & 1 & 1 \\
1 & 1 & 1 \\
1 & 1 & 1
\end{array} \right) \left(
\begin{array}{c} \overline{{\bf 5}}_1 \\ \overline{{\bf 5}}_2
\\ \overline{{\bf 5}}_3 \end{array}
\right) \frac{ \langle {\bf 5}_H \rangle \langle {\bf 5}_H \rangle}{M_R}.
\end{array}
\end{equation}

\noindent
This structure is characteristic of the kind of doubly lopsided models discussed
in Refs. \cite{bb1996, hm2001}.
This structure would give mass matrices for the up-type quarks, down-type quarks,
charged leptons, and neutrinos (denoted respectively by the subscripts
$U$, $D$, $L$, and $\nu$) of the form

\begin{equation}
\begin{array}{ll}
M_U \sim \left( \begin{array}{ccc}
\delta^2 \epsilon^2 & \delta \epsilon^2 & \delta \epsilon \\
\delta \epsilon^2 & \epsilon^2 & \epsilon \\
\delta \epsilon & \epsilon & 1
\end{array} \right) m, &  \\ & \\
M_D \sim \left( \begin{array}{ccc}
\delta \epsilon & \delta \epsilon & \delta \epsilon \\
\epsilon & \epsilon & \epsilon \\
1 & 1 & 1
\end{array} \right) m', &
M_L \sim \left( \begin{array}{ccc}
\delta \epsilon & \epsilon & 1 \\
\delta \epsilon & \epsilon & 1 \\
\delta \epsilon & \epsilon & 1
\end{array} \right) m' \\ & \\
M_{\nu} \sim \left( \begin{array}{ccc}
1 & 1 & 1 \\
1 & 1 & 1 \\
1 & 1 & 1
\end{array} \right) m_{\nu}. &
\end{array}
\end{equation}

\noindent From these forms several things are immediately apparent:
(a) the MNS neutrino mixing angles will be of order 1, (b) the CKM
quark mixing angles will be small (the 12 mixing of order $\delta$,
the 23 mixing of order $\epsilon$, and the 13 mixing of order
$\delta \epsilon$, (c) the masses of the up-type quarks will have a
strong family hierarchy $(\delta \epsilon)^2$: $(\epsilon)^2$: 1, (d) the
masses of the down-type quarks and charged leptons will have a weaker 
family hierarchy
$\delta \epsilon$: $\epsilon$: 1, and (e) the neutrino masses will
have the weakest family hierarchy, since all the neutrino masses will be of
roughly the same order. These five features are just exactly what is 
observed.

As we will see below, $SU(N)$ unification naturally leads to exactly the result
that the ${\bf 10}$'s of fermions are distinguished from each other by
symmetry --- symmetries in $SU(N)/SU(5)$ --- whereas the $\overline{{\bf 5}}$'s
of fermions are not distinguished by symmetry.

\section{An $SU(8)$ model: particle content}

We shall now describe a model based on $SU(8)$ where the
$SU(8)$ symmetry is sufficient to produce a non-trivial flavor structure
very much like that observed in nature.

If the number of left-handed
fermion multiplets of type $[p]$ and $\overline{[p]}$ is denoted by $n_p$
and $n_{-p}$ respectively, then the
condition that the $SU(8)$ anomalies cancel is $(n_1 - n_{-1}) + 4(n_2 - n_{-2})
+ 5(n_3 - n_{-3}) = 0$, and the condition for three families is
$(n_2 - n_{-2}) + 2(n_3 - n_{-3}) = 3$. The
general solution is $(n_1 - n_{-1}) = -12 +3p$, $(n_2 = n_{-2})
= 3 - 2p$, $(n_3 - n_{-3}) = p$.  The most economical set, as
measured by the total number of components, is $n_{-1} = 9$, $n_2 = 1$,
$n_3 = 1$, i.e. the set $[3] + [2] + 9 \times \overline{[1]} =
{\bf 56} + {\bf 28} + 9 \times \overline{{\bf 8}}$.
This is precisely the set of fermions that will be assumed in the
model presented below.

These fermion multiplets decompose under $SU(5)$ as follows.

\begin{equation}
\begin{array}{rcccccccccl}
\; [2]_L & =  & \psi^{[AB]} & \rightarrow & \psi^{\alpha \beta} & + & \psi^{\alpha I} & + & \psi^{IJ} & & \\
 &  & {\bf 28} & \rightarrow & {\bf 10} & + & 3 \times {\bf 5} & + & 3 \times {\bf 1}, & &  \\
& & & & & & & & & & \\
 \;[3]_L & = & \psi^{[ABC]} & \rightarrow & \psi^{\alpha \beta \gamma} & +
& \psi^{\alpha \beta I} & + &
\psi^{\alpha IJ} & + & \psi^{IJK} \\
& & {\bf 56} & \rightarrow & \overline{{\bf 10}} & + & 3 \times {\bf 10} & + &
3 \times {\bf 5} & + & {\bf 1}, \\
& & & & & & & & & & \\
\; 9 \times \overline{[1]}_L & = & \psi_{(m)A} & \rightarrow & \psi_{(m)\alpha} & + & \psi_{(m)I} & & & & \\
& & 9 \times \overline{{\bf 8}} & \rightarrow &
9 \times \overline{{\bf 5}} & + & 27 \times {\bf 1}, & & & & \\
\end{array}
\end{equation}

\noindent The subscripts $L$ on $[p]$ indicate that these are left-handed fermion multplets. The indices $A$, $B$, $C$, etc. run from 1 to 8; the
indices $\alpha$, $\beta$, $\gamma$ etc. run from 1 to 5; and the
indices $I$, $J$, $K$, etc. run from 6 to 8. All of the foregoing are
$SU(8)$ gauge indices.  The index $m = 1, ..., 9$, on the other hand, just
labels the nine
different antifundamental fermion multiplets.  One sees from Eq. (3) that there
are altogether four ${\bf 10}$ and one $\overline{{\bf 10}}$ of
$SU(5)$, for a ``net" of three ${\bf 10}$, and nine $\overline{{\bf
5}}$ and six ${\bf 5}$ of $SU(5)$, for a net of three
$\overline{{\bf 5}}$.  (It should be emphasized that we refer to
$SU(5)$ multiplets as a convenient way to keep track of the fermion
families, even though the actual sequence of breaking of $SU(N)$ to
the Standard Model group may not go through $SU(5)$. The sequence of
breaking depends on the relative magnitudes of the superlarge VEVs
of the model.) Which of the ${\bf 10}$ and which of the
$\overline{{\bf 5}}$ remain light after $SU(N)$ breaks to the
Standard Model depends on the Higgs content of the model, to which we now turn.

In the model it is assumed that the Higgs fields are in the following multiplets:
$[1]_H = H^A = {\bf 8}$, $[2]_H = H^{[AB]} = {\bf 28}$, $[4]_H = H^{[ABCD]} = {\bf 70}$,
and $Adj_H = \Omega^A_B =
{\bf 63}$. The $[1]_H$ and $[2]_H$ are
assumed to have superlarge VEVs in all the directions that leave the $SU(5)$ unbroken:
i.e. $H^I$ and $H^{IJ}$, $I,J = 6,7,8$. The $[4]_H$ has no
$SU(5)$-singlet components and so must not obtain a superlarge VEV.
The adjoint Higgs field has a superlarge diagonal VEV, which is needed for
the breaking to the Standard Model.
All three kinds of antisymmetric-tensor Higgs fields, $[1]_H$, $[2]_H$, and $[4]_H$,
participate in the breaking of $SU(2)_L \times U(1)_Y$ at the weak scale via
the weak doublets they contain, $H^i$, $H^{iI}$, and
$H^{iIJK}$, where $i = 1,2$. Of course, actually there is only one light Higgs doublet, which is a linear combination of these fields.

\section{Yukawa terms and superheavy fermion masses}

The renormalizable Yukawa terms that are allowed by $SU(8)$ are the following:

\begin{equation}
\begin{array}{ccl}
([3]_L \overline{[1]}_L) \; \overline{[2]}_H & = &
Y_m \; (\psi^{[ABC]} \; \psi_{(m)A}) \; H^*_{[BC]} \\
& & \\
([2]_L [2]_L) \; \overline{[4]}_H & = & Y \; (\psi^{[AB]} \; \psi^{[CD]}) \; H^*_{[ABCD]} \\
& & \\
([2]_L \overline{[1]}_L) \; \overline{[1]}_H & = & y_m \; (\psi^{[AB]} \; \psi_{(m)A}) \;
H^*_B \\
& & \\
( \overline{[1]}_L \overline{[1]}_L) \; [2]_H & = & a_{mn} \; (\psi_{(m)A} \; \psi_{(n)B}) \; H^{[AB]} \\
& & \\
\end{array}
\end{equation}

\noindent
A term of the form $([3]_L [3]_L) \; [2]_H$ vanishes by the antisymmetry of the tensors.
For the same reason, the Yukawa coupling matrix $a_{mn}$ in the fourth line of Eq. (4) is antisymmetric.
Note that $H^*_{[ABCD]} = \epsilon_{[ABCDEFGH]} \; H^{[EFGH]}/4! $.
Of course,
repeated indices of all kinds are summed over throughout this paper.

The first task is to determine how the vectorlike fermion pairs ``mate" to obtain 
superlarge mass, and which ones do, so as to identify the fermion multiplets that remain light. The ``mating" of the vectorlike pairs ${\bf 5} + \overline{{\bf 5}}$ that 
gives them superheavy masses is done by terms like
$y_m  ( \psi^{\alpha I} \; \psi_{(m) \alpha})  \langle H_I \rangle$ and
$Y_m ( \psi^{\alpha IJ} \; \psi_{(m) \alpha})  \langle H_{IJ} \rangle$.
It is clear that if there is only a single $[1]_H$ the former term mates only
one of the three ${\bf 5}$'s that are contained in the $[2]_L$, namely the linear combination
$\langle H_I \rangle \psi^{\alpha I}$. (It mates it with one of the
$\overline{{\bf 5}}$'s from among the nine $\overline{[1]}_L$,
namely the linear combination
$y_m \psi_{(m) \alpha}$.)  In order for
all three ${\bf 5}$'s that are contained
in the $[2]_L$ to be mated by renormalizable terms, there would have to be
three distinct $[1]_H$ multiplets. In that case, the mass term would be written
$y_{ma} (\psi^{\alpha I} \; \psi_{(m) \alpha}) \langle H_{(a) I} \rangle$,
$a = 1,2,3$, and for for each value of $a$ one ${\bf 5} + \overline{{\bf 5}}$ pair
would get mated. However, it is not necessary for the model to be complicated in that way.
Even with only a single $[1]_H$ of Higgs, all the ${\bf 5}$'s in the $[2]_L$ get mated if
higher-dimension operators induced by one-loop diagrams are taken into account.
For example, the one-loop diagrams shown in Fig. 1(a) and 1(b) induce the effective operators
$y_{m'} a_{m' m} (\psi^{\alpha I} \; \psi_{(m) \alpha} ) H^*_{IJ} H^J$ and
$y_{m^{\prime \prime}} a_{m^{\prime \prime} m'} a_{m' m}
(\psi^{\alpha I} \; \psi_{(m) \alpha} ) H^*_{IJ} \Omega^J_{J'} H^{J'}$.

\begin{picture}(360,180)
\thicklines
\put(60,60){\vector(1,0){30}}
\put(90,60){\line(1,0){30}}
\put(120,60){\line(1,0){30}}
\put(180,60){\vector(-1,0){30}}
\put(180,60){\vector(1,0){30}}
\put(210,60){\line(1,0){30}}
\put(240,60){\line(1,0){30}}
\put(300,60){\vector(-1,0){30}}
\put(180,100){\line(0,1){20}}
\put(170,125){$H^*_{IJ}$}
\put(88,90){$H^*_{\alpha IJK}$}
\put(240,90){$H^{\alpha K}$}
\put(80,45){$\psi^{\alpha I}$}
\put(140,45){$\psi^{JK}$}
\put(190,45){$\psi_{(m')K}$}
\put(260,45){$\psi_{(m)\alpha}$}
\put(180,40){\line(0,1){20}}
\put(170,25){$H^J$}
\put(120,53){$_Y$}
\put(235,55){$_{a_{m'm}}$}
\put(170,65){$_{y_{m'}}$}
\put(160,0){{\bf Fig. 1(a)}}
\put(180,60){\oval(120,80)[t]}
\end{picture}

\begin{picture}(360,180)
\thicklines
\put(60,60){\vector(1,0){30}}
\put(90,60){\line(1,0){30}}
\put(120,60){\line(1,0){30}}
\put(180,60){\vector(-1,0){30}}
\put(180,60){\vector(1,0){30}}
\put(210,60){\line(1,0){30}}
\put(240,60){\line(1,0){30}}
\put(300,60){\vector(-1,0){30}}
\put(180,100){\line(-1,1){15}}
\put(180,100){\line(1,1){15}}
\put(150,120){$H^{J'}$}
\put(200,120){$\Omega^J_{J'}$}
\put(100,90){$H^*_{\alpha}$}
\put(240,90){$H^{\alpha J}$}
\put(80,45){$\psi^{\alpha I}$}
\put(140,45){$\psi_{(m^{\prime \prime}) I}$}
\put(190,45){$\psi_{(m')J}$}
\put(260,45){$\psi_{(m)\alpha}$}
\put(180,40){\line(0,1){20}}
\put(170,25){$H_{IJ}$}
\put(110,55){$_{y_{m^{\prime \prime}}}$}
\put(230,55){$_{a_{m' m}}$}
\put(170,65){$_{a_{m^{\prime \prime} m'}}$}
\put(160,0){{\bf Fig. 1(b)}}
\put(180,60){\oval(120,80)[t]}
\end{picture}

\noindent
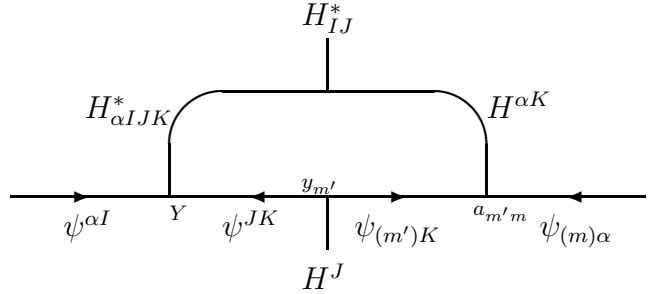
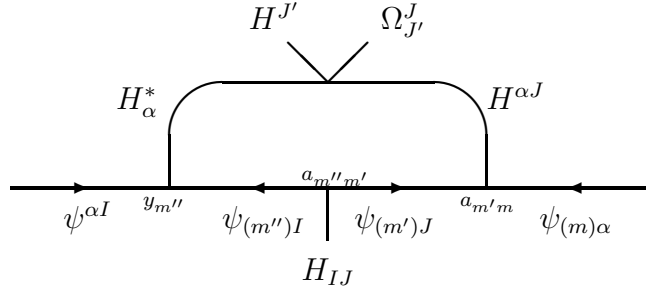
{\bf Figure 1:} Typical one-loop diagrams that ``mate" fermions in ${\bf 5}$ and
$\overline{{\bf 5}}$ multiplets of $SU(5)$ to give them superheavy mass.

\vspace{0.3cm}

In a similar way, if
there is only a single $[2]_H$ Higgs multiplet, the term
$y_m ( \psi^{\alpha IJ} \; \psi_{(m) \alpha})  \langle H_{IJ} \rangle$
only mates a single ${\bf 5}$ from the $[3]$ with a $\overline{{\bf 5}}$;
but loop diagrams induce higher-dimension
operators that mate the remaining two ${\bf 5}$'s from the $[3]_L$.
The mating of the $\overline{{\bf 10}}$ that is in the $[3]_L$ with a ${\bf 10}$
is not done by any renormalizable operator, but is done by such higher-dimension
operators as
$\epsilon_{\alpha \beta \gamma \delta \epsilon IJK} (\psi^{\alpha \beta \gamma}
\psi^{\delta \epsilon I'}) \Omega_{I'}^I H^{JK}$ and
$\epsilon_{\alpha \beta \gamma \delta \epsilon IJK} (\psi^{\alpha \beta \gamma}
\psi^{\delta \epsilon}) H^I H^{JK}$ . (The adjoint Higgs in the first operator
is needed to prevent it from vanishing identically by antisymmetry of indices.)
These operators
come from one-loop diagrams. They mate the $\overline{{\bf 10}}$ with some linear
combination of the ${\bf 10}$'s from the $[3]_L$ and $[2]_L$.

\section{The light families and their masses}

One sees, then, that even the small set of Higgs multiplets given above,
$H^A$, $H^{[AB]}$, $H^{[ABCD]}$, and $\Omega^A_B$, with one of each type, is enough
to mate all of the conjugate pairs of fermion multiplets and make them superheavy.
Which fermion multiplets mate determines which multiplets remain light.

The three ${\bf 10}$'s that remain light are linear combinations of
the one that is in $[2]_L$ and the three that are in $[3]_L$.
Without loss of generality, we can choose the
flavor basis of the light fermions so that ${\bf 10}_3$ comes partly from
$[2]_L$, but that ${\bf 10}_1$ and ${\bf 10}_2$ come purely from $[3]_L$.
This shows that for the ${\bf 10}$'s one family is automatically selected out
as different by virtue of coming partly from a different $SU(8)$ multiplet than the other
families. This will allow an explanation of why the $t$ quark
is so much heavier than the $u$ and $c$ quarks. Moreover, even though
the ${\bf 10}_1$ and ${\bf 10}_2$ come entirely from the same $SU(8)$ multiplet, namely
$[3]_L$, they come from different {\it components} of that multiplet.  That is,
they are given by $\psi^{\alpha \beta I}$ with different values of the
$SU(8)/SU(5)$ index $I$ and are thus distinguished from each other by SU(8).
Thus, $SU(8)$ can suppress the mixing of these ${\bf 10}$'s, as will be seen.

By contrast, one sees that all three light
$\overline{{\bf 5}}$'s must come from the same kind of $SU(8)$ multiplet, namely
$\overline{[1]}_L$. In other words, the three light $\overline{{\bf 5}}$'s are
simply three particular linear combinations of the nine $\psi_{(m) \alpha}$. (For simplicity,
we could take the basis in the space of these nine fields to
be such that the light ones corresponded to the values $m = 1,2,3$.)
Since $\psi_{(m) \alpha}$ has only an $SU(5)$ index and a label $(m)$ that has
nothing to do with the gauge symmetry, the $SU(8)$ does not
distinguish among the three light $\overline{{\bf 5}}$'s in any way. One would
therefore expect that these
$\overline{{\bf 5}}$'s would be able to mix strongly with each other.  

It is interesting that the large mixing
among $\overline{{\bf 5}}$'s that is an ingredient of the lopsided and
doubly lopsided models emerges naturally in the context of
$SU(N)$ unification with $N>5$.  The reason has to do with anomaly
cancellation. The ${\bf 10}$'s of $SU(5)$ must come from tensors
that have a rank of at least 2, which tend (for large $N$) to make a
large positive contribution to the anomaly. In the most economical
solutions of the anomaly conditions, this large contribution tends
to be cancelled by large numbers of antifundamental multiplets. This,
in turn, gives the result in many cases that the light
$\overline{{\bf 5}}$'s all come from antifundamentals, as in the
present $SU(8)$ example. To take another example, in $SU(9)$ the most
economical three-family solutions to the anomaly conditions are (a)
$[3] + 9 \times \overline{[1]}$ (165 components) and (b) $3 \times [2]
+ 15 \times \overline{[1]}$ (243 components). Both of these solutions have numerous
antifundamentals, and in both solutions all of the $\overline{{\bf 5}}$ are contained in
these antifundamentals.

The masses of the up-type quarks, $u$, $c$, and $t$, come from
operators that (in $SU(5)$ terms) couple ${\bf 10}_L$ to ${\bf 10}_L$.
There is only one renormalizable operator of this type, namely

\begin{equation}
{\cal O}_A = ([2]_L [2]_L) \overline{[4]}_H =
\psi^{AB} \psi^{CD} H^*_{ABCD},
\end{equation}

\noindent
which contains the term $(\psi^{\alpha \beta} \psi^{\gamma \delta}) H^*_{\alpha
\beta \gamma \delta}$. (Note that $H^*_{\alpha
\beta \gamma \delta} = \epsilon_{\alpha \beta \gamma \delta \epsilon 678} H^{\epsilon 678}$.)
However, only one of the light ${\bf 10}_L$'s, namely the one that we have labelled
${\bf 10}_3$, contains some of $[2]_L$, i.e. of $\psi^{\alpha \beta}$; the other two light
${\bf 10}$'s, namely ${\bf 10}_1$ and ${\bf 10}_2$, are purely in $[3]_L$.
Consequently the operator ${\cal O}_A$ contributes only to the 33 element of $M_U$,
the mass matrix of the up-type quarks. This element, which will be denoted $A$, is
the only element of $M_U$ that arises at tree level, thus explaining the relatively large magnitude of the $t$-quark mass.

At one-loop level, however, many higher-dimension
operators are induced that contribute to the other elements of $M_U$. In particular, one
has the following classes of operators:

\begin{equation}
\begin{array}{ccl}
{\cal O}_{\beta} & = & ([2]_L [3]_L) [1]_H [2]_H, \;\; ([2]_L[3]_L) \overline{[1]}_H [4]_H, ... \\
& = & \epsilon_{ABCDEFGH} (\psi^{AB} \psi^{CDE}) H^F H^{GH}, \;\;
\epsilon_{ABCDEFGH} (\psi^{AB} \psi^{CDI}) H_I H^{EFGH}, ... \\
\\& & \\
{\cal O}_{\gamma} & = & ([3]_L [3]_L) Adj_H [2]_H, \;\;
([3]_L [3]_L) \overline{[2]}_H [4]_H, ... \\
& = & \epsilon_{ABCDEFGH}
(\psi^{ABC} \psi^{DEI}) \Omega^F_I H^{GH}, \;\;
\epsilon_{ABCDEFGH}
(\psi^{ABC} \psi^{DEI}) H_{IJ} H^{JFGH}, ... \\
\\ & & \\
{\cal O}_{\delta} & = & ([3]_L [3]_L) \overline{[1]}_H [1]_H [2]_H, \;\;
([3]_L[3]_L) \overline{[1]}_H \overline{[1]}_H [4]_H, ... \\
& = & \epsilon_{ABCDEFGH} (\psi^{ABC} \psi^{DEI}) H_I H^F H^{GH},
\;\;
\epsilon_{ABCDEFGH} (\psi^{ABI} \psi^{CDJ}) H_I H_J H^{EFGH}, ...
\end{array}
\end{equation}

\noindent
The operators of type ${\cal O}_{\beta}$ couple $[2]_L$ to $[3]_L$, and therefore
couple ${\bf 10}_3$ to ${\bf 10}_1$ and ${\bf 10}_2$. These operators thus contribute to
the 13 (31) and 23 (32) elements of $M_U$, which will be denoted $\beta'$ and
$\beta$, respectively. (The operators ${\cal O}_{\beta}$ will also contribute to
the 33 element $A$.)

The operators of type ${\cal O}_{\gamma}$ couple $[3]_L$ to $[3]_L$, and therefore
couple any of the ${\bf 10}_i$ to any {\it other} of the ${\bf 10}_i$. They cannot,
however, contribute to any diagonal element of $M_U$, because of the antisymmetry of the
epsilon symbol.  These operators therefore contribute to the 12 (21) element of
$M_U$, which is denoted $\gamma$, as well as to the elements $\beta, \beta'$.

Finally, operators of the type ${\cal O}_{\delta}$, which also couple $[3]_L$ to
$[3]_L$, can contribute to any elements of $M_U$, including the 11 and 22 elements,
which are denoted $\delta'$ and $\delta$, respectively.

In sum, the mass matrix of the up-type quarks has the form

\begin{equation}
M_U = \left( \begin{array}{ccc}
\delta' & \gamma & \beta' \\
\gamma & \delta & \beta \\
\beta' & \beta & A
\end{array}
\right)
\end{equation}

There is no reason {\it a priori} why the different types of operators induced at
one-loop level must all make contributions to $M_U$ of the same order of
magnitude. For example,
the operators of type ${\cal O}_{\delta}$ are of dimension 6 or higher, whereas some of
the operators of type ${\cal O}_{\beta}$ are only of dimension 5. So it could be that
$\delta, \delta' \ll \beta, \beta'$. Moreover, the superheavy
VEVs of Higgs fields in different
representations of $SU(8)$ could be of quite different magnitudes, so that
even operators of the same dimension but involving different types of
Higgs multiplets could make very different contributions.

If it were the case that $\gamma, \delta, \delta' \ll \beta, \beta'$, then the matrix $M_U$ would have the observed threefold hierarchy among its eigenvalues, i.e. $m_u \ll m_c \ll m_t$.

Turning now to the masses of the down-type quarks and charged leptons, these come from
operators that (in $SU(5)$ terms) couple ${\bf 10}_L$ to $\overline{{\bf 5}}_L$.
At first glance, there seem to be dimension-4 operators that do this, namely

\begin{equation}
\begin{array}{l}
y_m \; (\psi^{\alpha \beta} \; \psi_{(m) \alpha}) \; H^*_{\beta},   \\
Y_m \; (\psi^{\alpha \beta I} \; \psi_{(m) \alpha}) \; H^*_{\beta I}.
\end{array}
\end{equation}

\noindent
However, the first of these operators is related by
$SU(8)$ to the operator $y_m \; (\psi^{\alpha I} \; \psi_{(m)
\alpha}) \; H^*_I$, which mates precisely the $\overline{{\bf 5}}_L$
that is the linear combination $y_m \psi_{(m) \alpha}$ to a ${\bf
5}$ to make it superheavy. So that the first term in Eq. (8) is not
a contribution to the light fermion mass matrices, but is a coupling
of light fermions to superheavy fermions. In the same way, the
second operator in Eq. (8) is related by $SU(8)$ to the operator
$Y_m \; (\psi^{\alpha IJ} \; \psi_{(m) \alpha}) \; H^*_{IJ}$, which
mates precisely the $\overline{{\bf 5}}_L$ that is the linear
combination $Y_m \psi_{(m) \alpha}$ to a ${\bf 5}$ to make it
superheavy. The second term in Eq. (8) is thus also not a
contribution to the mass matrices of the light fermions.

The mass matrices of the down-type quarks and charged leptons, which will be denoted
$M_D$ and $M_L$, respectively, do not arise until one-loop.  There are two kinds of
operators that contribute:

\begin{equation}
\begin{array}{ccl}
{\cal O}_{\epsilon} & = & ([2]_L \overline{[1]}_L) Adj_H \overline{[1]}_H,
\;\; ([2]_L \overline{[1]}_L) \overline{[2]}_H [1]_H, ... \\
& = & (\psi^{AB'} \psi_{(m)A}) \Omega^B_{B'} H_B, \;\;
(\psi^{AB'} \psi_{(m)A} ) H_{B'C} H^C, ... \\
\\& & \\
{\cal O}_{\zeta} & = & ([3]_L \overline{[1]}_L) \overline{[1]}_H Adj_H \overline{[1]}_H, \;\;
([3]_L \overline{[1]}_L) Adj_H \overline{[2]}_H, ... \\
& = &
(\psi^{ABC'} \psi_{(m)A}) H_B \Omega_{C'}^C H_C, \;\;
(\psi^{ABC'} \psi_{(m)A}) \Omega_{C'}^C H_{BC}, ...
\end{array}
\end{equation}

\noindent
The operators of type ${\cal O}_{\epsilon}$ couple $[2]_L$ to $\overline{[1]}$ and
therefore ${\bf 10}_3$ to $\overline{{\bf 5}}_i$, $i=1,2,3$. Thus they contribute to the
$3i$ elements of $M_D$ and the $i3$ elements of $M_L$, which we denote
$\epsilon_i$.
The operators of type ${\cal O}_{\zeta}$ couple $[3]_L$ to $\overline{[1]}$ and therefore
can contribute to all the elements of the mass matrices $M_D$ and $M_L$. We denote the
resulting non-vanishing $2i$ elements of $M_D$ and $i2$ elements of $M_L$ by
$\zeta_i$, and the resulting non-vanishing $1i$ elements of $M_D$ and $i1$ elements of
$M_L$ by $\zeta'_i$. These matrices consequently have the form,

\begin{equation}
M_D = \left( \begin{array}{ccc}
\zeta'_1 & \zeta'_2 & \zeta'_3 \\
\zeta_1 & \zeta_2 & \zeta_3 \\
\epsilon_1 & \epsilon_2 & \epsilon_3
\end{array} \right), \;\;\;
M_L \sim \left( \begin{array}{ccc}
\zeta'_1 & \zeta_1 & \epsilon_1 \\
\zeta'_2 & \zeta_2 & \epsilon_2 \\
\zeta'_3 & \zeta_3 & \epsilon_3
\end{array} \right).
\end{equation}

\noindent
The matrix $M_L$ is not exactly the transpose of $M_D$,
because of $SU(5)$-breaking effects from the adjoint Higgs VEVs that
come into the one-loop diagrams (e.g. the factors of $\Omega^B_{B'}$ in
Eq. (9)). That is why a ``$\sim$" is used in the equation for
$M_L$ rather than an equal sign. These $SU(5)$-breaking effects can
explain the well-known Georgi-Jarlskog factors \cite{gj}, i.e. the deviations
of $m_s/m_{\mu}$ and $m_d/m_e$ from 1.

The notation used in writing elements of the mass matrices is as follows:

(a) Elements
that come from operators of the same class are denoted by the same Greek letter. For example, $\beta$ and $\beta'$ in Eq. (7) both come from the operators of class
${\cal O}_{\beta}$, and $\zeta_1$, $\zeta_2$, $\zeta_3$, $\zeta'_1$, $\zeta'_2$,
and $\zeta'_3$ all come from the operators of class ${\cal O}_{\zeta}$. Consequently,
elements that are denoted by different Greek letters, since they come from entirely {\it different operators}, have no reason to be comparable in magnitude.

(b) Elements that are denoted by the same Greek letter but differ by a prime,
such as $\beta$ and $\beta'$ or
$\zeta_i$ and $\zeta'_i$, come from the same operators, containing the same
$SU(8)$ multiplets, but involve {\it different components} of those multiplets.
For example, suppose that ${\bf 10}_1 = \psi^{\alpha \beta 8}$ and
${\bf 10}_2 = \psi^{\alpha \beta 7}$. Then the elements $\beta$ and $\beta'$
would both come from the operators ${\cal O}_{\beta}$, but $\beta$ would come from
the terms $(\psi^{\alpha \beta} \psi^{\gamma \delta 7}) H^{\epsilon} H^{86}$,
$(\psi^{\alpha \beta} \psi^{\gamma \delta 7}) H_7 H^{\epsilon 678}$, etc., whereas
$\beta'$ would come from the terms
$(\psi^{\alpha \beta} \psi^{\gamma \delta 8}) H^{\epsilon} H^{67}$,
$(\psi^{\alpha \beta} \psi^{\gamma \delta 8}) H_8 H^{\epsilon 678}$, etc..
Since different components of the same $SU(8)$ multiplet of Higgs fields
--- such as $H^6$, $H^7$, and $H^8$ --- can have vacuum expectation values that are
very different from each other if there is a hierarchy of scales involved
in the breaking of $SU(8)$ down to the Standard Model group, elements that differ by
a prime can also be of very different magnitude.
In other words, we see that a hierarchy among elements of a mass matrix of light fermions, i.e. a ``flavor hierarchy", can arise in part from a hierarchy of scales in
the breaking of the grand unified group.

(c) Elements that are distinguished only by a subscript, such as
$\zeta'_2$ and $\zeta'_3$, come from the same kinds of operators,
and the same $SU(8)$ components of the multiplets within those
operators, but involve {\it different antifundamental multiplets of
fermions}. For example, $\epsilon_1$, $\epsilon_2$, and $\epsilon_3$
all come from the same operators ${\cal O}_{\epsilon}$ (such as
$\psi^{AB'} \psi_{(m) A} \Omega^B_{B'} H_B$) and with the SU(8)
indices taking the same values; but they involve different linear
combinations of the nine antifundamental multiplets $\psi_{(m) A}$,
$m = 1, ..., 9$. In other words, $SU(8)$ gauge symmetry in no way
distinguishes among the elements $\epsilon_1$, $\epsilon_2$, and
$\epsilon_3$. If there are no preferred directions in the
nine-dimensional space spanned by the index $m$ --- i.e. if the
Yukawa couplings $Y_m$, $y_m$, and $a_{mn}$ are ``randomly" oriented
in that space --- then one expects that $\epsilon_1 \sim \epsilon_2
\sim \epsilon_3$, $\zeta_1 \sim \zeta_2 \sim \zeta_3$, and $\zeta'_1
\sim \zeta'_2 \sim \zeta'_3$.

In consequence, one expects the matrices $M_D$ and $M_L$ to have a
{\it stratified} structure characteristic of the doubly lopsided models of
Refs. \cite{bb1996, hm2001}. All the elements of a row of $M_D$ (or a
column of $M_L$) should be comparable in magnitude; whereas the
different rows of $M_D$ (or columns of $M_L$) should typically be
quite different in magnitude. As was explained in the Introduction, such a stratified
structure leads to a
situation where the mixing angles of the left-handed quarks (the CKM
angles) are small, while the mixing angles of the left-handed
leptons (the MNS neutrino-mixing angles) are of order one. This is
clear from a direct inspection of the mass matrices: the CKM angles
evidently involve ratios of elements of different rows of $M_D$
(e.g. $V_{cb}$ would involve $\zeta_3/\epsilon_3 \ll 1$), while the MNS
angles involve elements of different rows of $M_L$ (e.g. $U_{\mu 3}
= \sin \theta_{atm}$ involves the ratio $\epsilon_2/\epsilon_3 \sim
1$).

Turning to the mass matrix of the light neutrinos, it is apparent
that all of its elements should be comparable, since the three light
neutrinos are not distinguished in any way by $SU(8)$, but only by
which antifundamental fermion multiplets they are contained in. That
is, they all come from the same kind of multiplets, $\psi_{(m) i}$. 
This would imply that the ratios of
neutrino masses should not exhibit a large hierarchy, which is
consistent with the fact that $(\Delta m^2_{sol})^{1/2}$ and
$(\Delta m^2_{atm})^{1/2}$ only differ by about a factor of 5. Since
each of the matrices $M_D$ and $M_L$ contains elements of various types,
(though they all arise at one-loop level) one expects a much
stronger hierarchy among their eigenvalues, as is indeed observed.
And finally, since $M_U$ not only contains elements of different
types, but also both tree-level and one-loop elements, the hierarchy
among the up-type quarks should be the strongest of all; and that
too corresponds to what is seen.

Finally, it should be noted that there are many Standard Model singlet fermion fields in this model, which can play the role of right-handed neutrinos. To be exact, there are 31 of them, of which 27 come from the nine antifundamental multiplets of $SU(8)$. For these 27, the masses come predominantly from the coupling $a_{mn} \psi_{(m) I} \psi_{(n) J}
\langle H^{IJ} \rangle$. Due to the antisymmetry of the matrix $a_{mn}$, these terms by themselves would lead to Dirac masses for these particles. When other contributions to the right-handed neutrino masses are taken into account, a ``pseudo-Dirac" form can emerge.  As is well-known, such a pseudo-Dirac structure can lead to resonant enhancement of leptogenesis.

\section{Conclusions}

It has been shown that a realistic grand unified model can be constructed based on $SU(N)$,
$N>5$, in which the $SU(N)$ symmetry and its pattern of breaking  is sufficient to create a non-trivial flavor structure for the light quarks and leptons, without there being any flavor symmetry at all. What makes the fermions of different families different from each other is the way they transform under the $SU(N)$. This is in particular true of the three light ${\bf 10}$'s of $SU(5)$, which do not all come from the same kinds of multiplets of $SU(N)$.
On the other hand, in this model the three light $\overline{{\bf 5}}$'s of $SU(5)$ do all come from the same kind of multiplet of $SU(N)$, and thus are not distinguished from each other. Since the left-handed neutrinos are all contained in the $\overline{{\bf 5}}$'s,
no fundamental symmetry distinguishes the light neutrinos from each other, and as a consequence large neutrino mixing naturally results and the neutrino masses should not exhibit a strong hierarchy. For the mass matrices of the down-type quarks and the charged leptons a stratified or ``doubly lopsided" structure results, leading to a stronger hierarchy for their masses.  The strongest mass hierarchy of all is that of the up-type quarks. (In the $SU(8)$ model we present as
an example, only the top quark obtains mass at tree level.)

The fact that the three light $\overline{{\bf 5}}$'s are not
distinguished by any symmetry (which is what gives the realistic
stratified structure to the mass matrices) stems from the fact that
they all come from antifundamental multiplets of $SU(N)$. That in
turn can be traced to the requirements of anomaly cancellation. For
$SU(N)$ models containing only antisymmetric tensor multiplets of
fermions, the most economical sets of fermions that have three
families and are anomaly free tend to have many antifundamental
multiplets and it is usually the case that all of the
$\overline{{\bf 5}}$'s come from these multiplets.

The model described above is a non-supersymmetric grand unified
theory. It is also possible to construct models based on the same
ideas that have low-energy supersymmetry. In such models all the
masses of the light families would have to come from tree-level
diagrams. However, there could still be mass hierarchies, since tree
diagrams can generate operators of different dimensions and of
different types. Moreover, there can be a hierarchy among the scales
at which $SU(N)$ breaks down to the Standard Model group, and this
hierarchy can be reflected in the mass matrices of the light quarks
and leptons, as the model presented here illustrates.

\end{document}